\documentclass[letter,twocolumn]{jpsj2}
\def\kf{k_{\rm F}}

\def\ggs{\buildrel\textstyle > \over {\hbox{\raise0.2ex\hbox{$\sim$}}}}
\def\lls{\buildrel\textstyle < \over {\hbox{\raise0.2ex\hbox{$\sim$}}}}
\def\gsim{\,\lower0.75ex\hbox{$\ggs$}\,}
\def\lsim{\,\lower0.75ex\hbox{$\lls$}\,}

\def\tt{\tilde{t}}

\def\ie{{\it i.e.}, }

\def\jo #1#2#3#4{#1 {\bf #2} (#3) #4}   
\def\PR{Phys.\ Rev.}
\def\PRB{Phys.\ Rev.\ B}
\def\PRL{Phys.\ Rev.\ Lett.}

\def\JPSJ{J.\ Phys.\ Soc.\ Jpn.}

\def\SOV{Sov.\ Phys.\ JETP}

\def\IJMPB{Int.\ J.\ Mod.\ Phys.\ B}

\title{Theoretical Study on Transport Properties of Normal Metal - Zigzag Graphene Nanoribbon -
Normal Metal Junctions}

\author{Yoneko \textsc{Mochizuki} and  
Hideo \textsc{Yoshioka}\thanks{E-mail address: h-yoshi@cc.nara-wu.ac.jp}}

\inst{Department of Physics, Nara Women's University, Nara 630-8506}

\recdate{\today}

\abst{We investigate transport properties of the junctions in which   
the graphene nanoribbon with the zigzag shaped edges 
consisting of the $N$ legs 
is sandwiched by the two normal metals by means of recursive Green's
function method. 
The conductance and the transmission probabilities  
are found to have the remarkable properties 
depending on the parity of $N$. 
The singular behaviors close to $E=0$ with $E$ being the Fermi energy 
are demonstrated.  
The channel filtering is shown to occur in the case with $N=$ even.  
}

\kword{graphene nanoribbon, zigzag shaped edges, transport property}

\begin{document}
\maketitle


Recently, graphene-based materials with nano-meter sizes have been
attracting much attention in both fundamental and applied sciences. 
Among these materials, the graphene nanoribbon with zigzag shaped edges, 
which is abbreviated to zigzag GNR in the following, has fascinating peculiar
properties as follows.\cite{Fujita,Nakada,Wakabayashi} 
The zigzag GNR has a metallic band structure irrespective of the width
$N$ in the sense that the energy gap does not appear at the Fermi
energy in the absence of doping, $E=0$. 
However, the asymptotic form of the energy dispersion near $E=0$ 
is written as $E \propto \pm |k-\pi/a|^N$ with $a$ being the lattice spacing 
and then Drude weight in the absence of doping becomes zero due to 
the vanishing Fermi velocity for $N \geq 2$.    
Therefore, according to Kohn's criterion,\cite{Kohn} 
the system without doping is classified into the insulator 
from the point of view for transport properties.
Such characteristic properties are due to the fact that 
the one-particle states near $E=0$ are localized  
around zigzag edges.      

Transport properties of the zigzag GNR applied to the external
potential have been studied
theoretically,\cite{Wakabayashi-Aoki,Akhmerov,Cresti,Nakabayashi} 
and it has been found that 
the parity of the width $N$ remarkably affects the transport properties.
In the case of $N=$ even, the zigzag GNR has the reflection symmetry 
in the transverse direction, 
and as a result, the parity of the wave function in the direction must be 
even or odd. 
Since only the scattering processes conserving the parity of the wave function 
are allowed under the external potential not varying in the
transverse direction, the filtering of the scattering processes happens
depending on the parity of the wave function of the transverse
direction.\cite{Nakabayashi}     
Note that such an interesting phenomenon depending on the parity of the
width $N$ have been found in the persistent current of the isolated ring
pierced by the magnetic flux.\cite{Yoshioka-Higashibata}  
However, the transport properties of the junctions which consist of the zigzag GNR and the normal metals have not been investigated. 

\begin{figure}[htb]
\begin{center}
\includegraphics[width=8.5truecm]{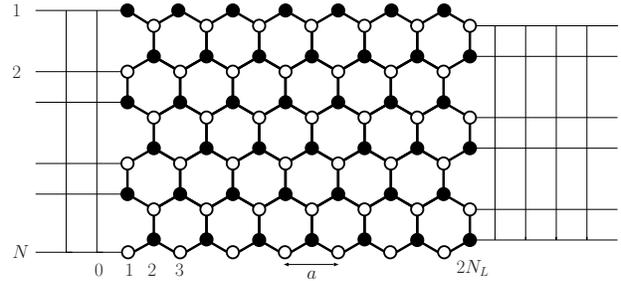}
\end{center}
\caption{
The present model where the zigzag GNR written by the thick lines 
is sandwiched by the ideal leads expressed by the regular square
 lattices, which is expressed by the thin lines.  
Here, the width and length are given by $N$ and $2N_L$, respectively,
 and $a$ is the lattice spacing of the zigzag GNR.  
The filled (open) circles in the zigzag GNR region express 
the A (B) sublattice.
}
\label{fig:model}
\end{figure}

In the present paper, we investigate the transport properties of the
junctions shown Fig. \ref{fig:model} where the zigzag GNR with the width $N$ and the
length $2 N_L$ is sandwiched by the two ideal leads expressed by the
regular square lattices.
The Hamiltonian is written as follows,  
\begin{equation}
 H = \sum_{i,j} t_{i,j} |i \rangle \langle j|, 
\end{equation}
where $|i \rangle$ is the localized state at the site $i$ 
and $t_{i,j} = -t $ if $i$ and $j$ are nearest neighbors, otherwise 
$t_{i,j} = 0$.  
The hopping in the GNR region is assumed to have the same value $t$ as that in
the normal metals for simplicity.  
The conductance in unit of $2e^2/h$ ( $-e < 0$ : electronic charge, $h$
: Planck constant), $g$, 
written by Landauer formula  
\begin{equation}
 g = \sum_{\mu, \nu} T_{\mu, \nu}
\end{equation}
is calculated by the recursive Green's function method.\cite{Ando}   
Here, $T_{\mu,\nu}$ is the transmission probability 
from the incident channel $\nu$ to transmitted one $\mu$, 
both of which are defined in the ideal leads. 
Note that we have confirmed that the conservation law 
$\sum_{\mu} \left\{ T_{\mu,\nu}+ R_{\mu,\nu} \right\} = 1$
holds in all the numerical calculation 
in order to check validity of our results 
where $R_{\mu,\nu}$ is the reflection probability from the channel $\nu$ to $\mu$.  

\begin{figure}[htb]
\begin{center}
\includegraphics[width=7.5truecm]{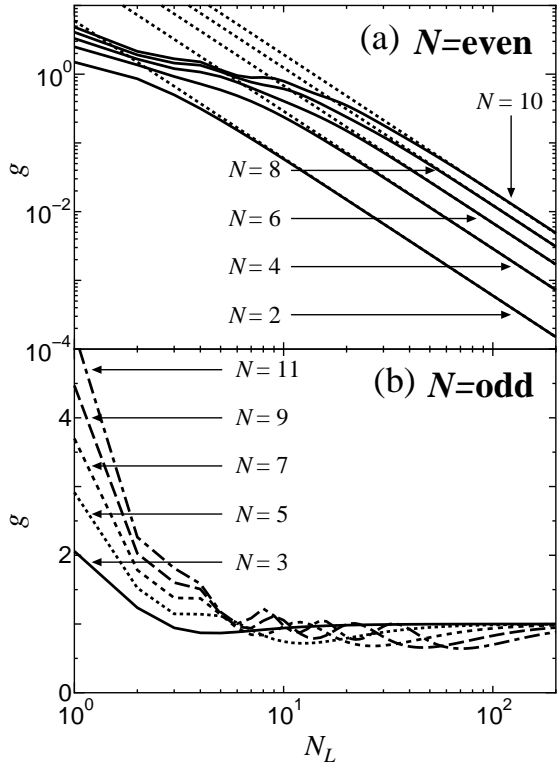} 
\end{center}
\caption{
The conductance $g$ at $E/t = 0$ as a function of $N_L$ for $N=$ 2, 4,
 6, 8, 10 (a) and for $N=$ 3, 5, 7, 9, 11 (b).  
In (a), the dotted lines express the asymptotic behaviors for $N_L \gg
 1$, all of which are proportional to $N_L^{-2}$.  
}
\label{fig:E=0}
\end{figure}

At first, the cases without doping, i.e., $E/t=0$ are discussed.\cite{Note}  
We show the conductance $g$ as a function of the length $N_L$ for 
$N=$ even in Fig. \ref{fig:E=0} (a) and $N=$ odd in Fig. \ref{fig:E=0}
(b), respectively.    
We can observe obvious qualitative discrepancy between $N=$ odd and $N=$ even 
in the asymptotic behavior for $N_L \gg 1$.  
In the case with $N=$ odd, the conductances approach to unity, 
which indicates that only the one channel survives in the limit.   
On the other hand, those for $N=$ even decrease in proportion to $N_L^{-2}$. 
Actually, we can successfully derive the expression 
$g\simeq 6/N_L^2$ by obtaining the recursive Green's function analytically
in the case of $N=2$.   
Note that the transmission probabilities are obtained as 
$T_{1,1} = T_{2,2} \simeq 3/N_L^2 $ and $T_{1,2}=T_{2,1} \simeq 0$ in
this limit. 

\begin{figure}[htb]
\begin{center}
\includegraphics[height=11.0truecm]{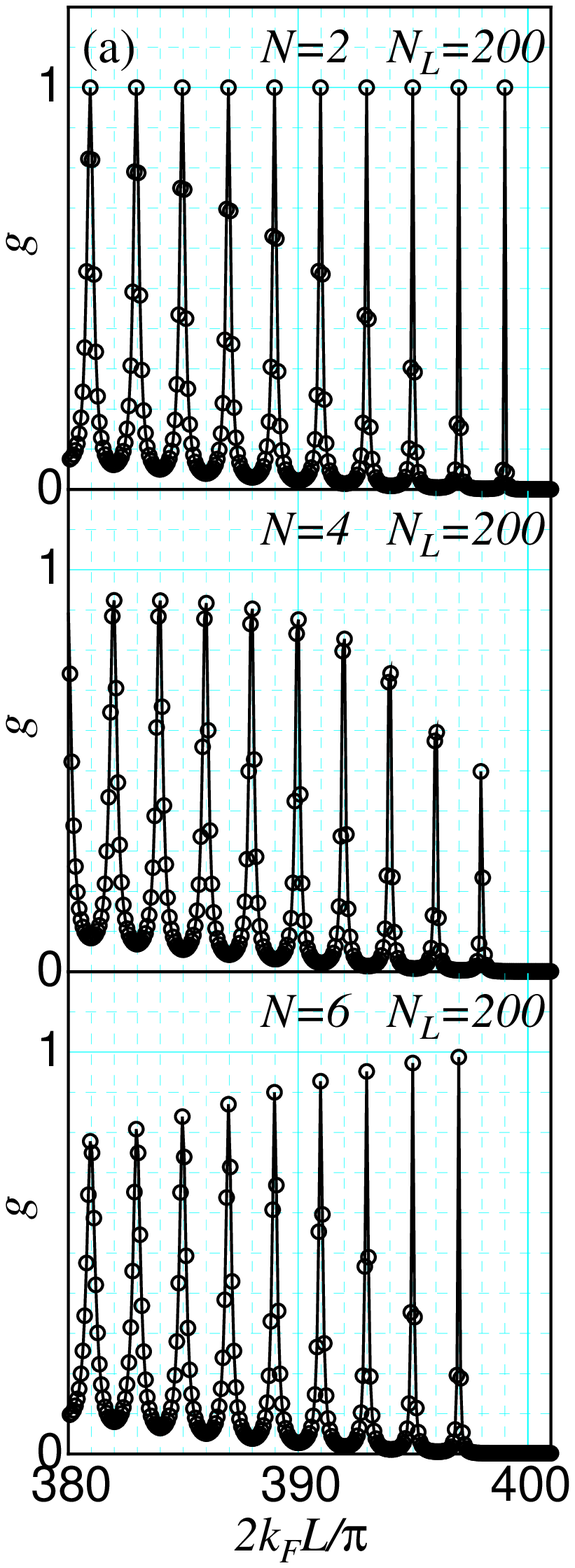} 
\includegraphics[height=11.0truecm]{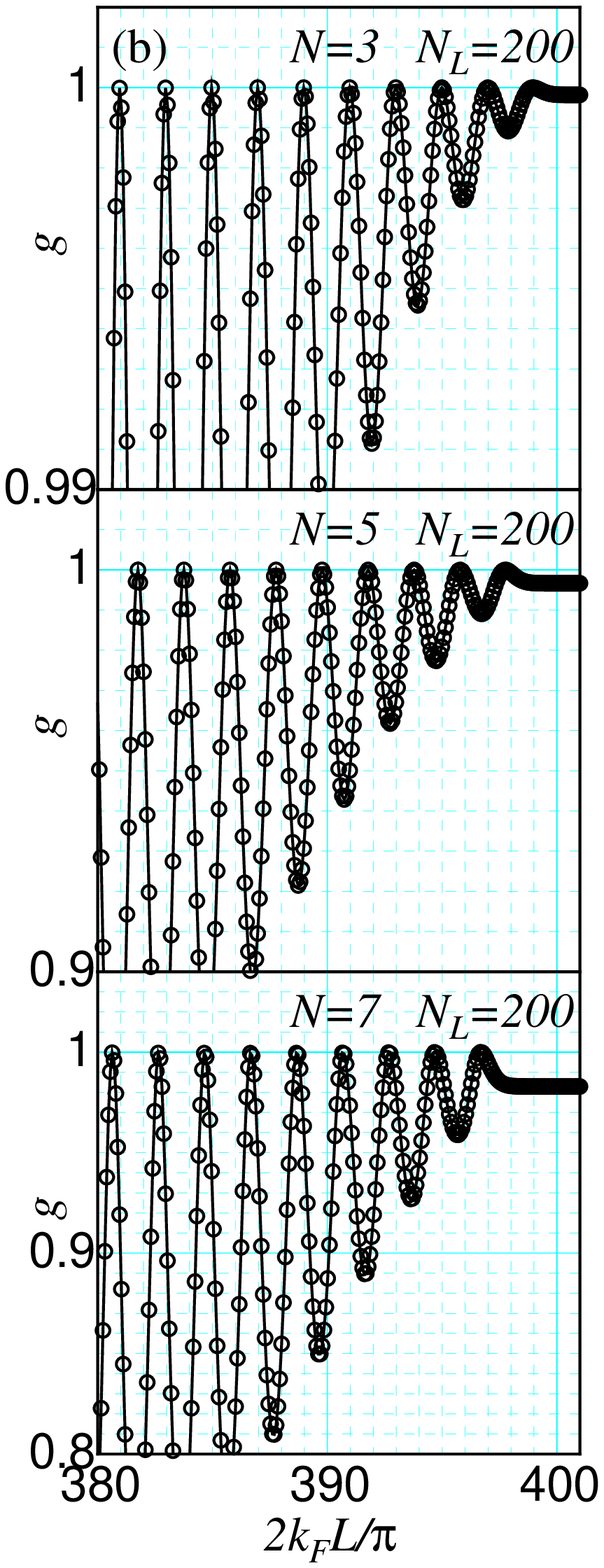} 
\end{center}
\caption{ (Color online)
The conductance $g$ at $E/t > 0$ as a function of $2 \kf L/\pi$ for $N=$
 even (a) and for $N=$ odd (b) with $L = (N_L+1/2)a$ being  
the length of the zigzag GNR region and $N_L = 200$.   
In each figure, the right end of horizontal axis $2 \kf L/\pi = 401$
corresponds to the $E/t=0$ case.     
}
\label{fig:Eneq0}
\end{figure}

Next, the cases with $E/t \neq 0$ are investigated.  
Due to the particle-hole symmetry, 
the transport properties for $E/t > 0$ are the same as 
those for $E/t < 0$. 
Therefore we concentrate on only the $E/t > 0$ cases unless explicitly noted.  
In addition, we discuss the energy regions where the one band 
gets across the Fermi energy. 
Here, there exits the one-to-one correspondence between 
the energy and the Fermi wavenumber $\kf$ ; 
$\kf$ is equal to $\pi/a$ at $E/t=0$ 
and decreases with increasing the energy.   
Therefore, the conductances are investigated as a function of the Fermi
wavenumber instead of the energy. 
In Fig. \ref{fig:Eneq0}, 
we show the conductances as a function of $2\kf L/\pi$ 
for $N$ = even (a) and $N$ = odd (b) 
 with the fixed length $N_L = 200$. 
In this analysis, we use $L = (N_L + 1/2) a$ as 
the length of the zigzag GNR region $L$. 
In each figure, the right end of the horizontal axis is given by $2 \kf L/\pi
= 401$ and it corresponds to $E/t=0$, \ie $k_F = \pi/a$. 
We can see the oscillating behavior in the both cases except near
$E/t=0$. 
The interval between one peak and the nearest neighbor one is given by
$\Delta (\kf L) = \pi$. 
Therefore, the oscillation is considered to be originated from the interference between 
the electron waves at the Fermi energy.  
It should be noted that such oscillating behavior disappears near $E/t=0$. 
In the cases of $N$=2 and 3 (4 and 5), 
the peak which should appear at $2 \kf L/\pi
= 401 (400)$ does not exist. 
With increasing the width of the zigzag GNR, 
the number of missing peaks increase. 
For example, the two peaks expected at $2 \kf L/\pi = 399$ and 401
respectively are missing in the cases of $N = 6$ and 7.

\begin{figure}[htb]
\begin{center}
\includegraphics[width=7.5truecm]{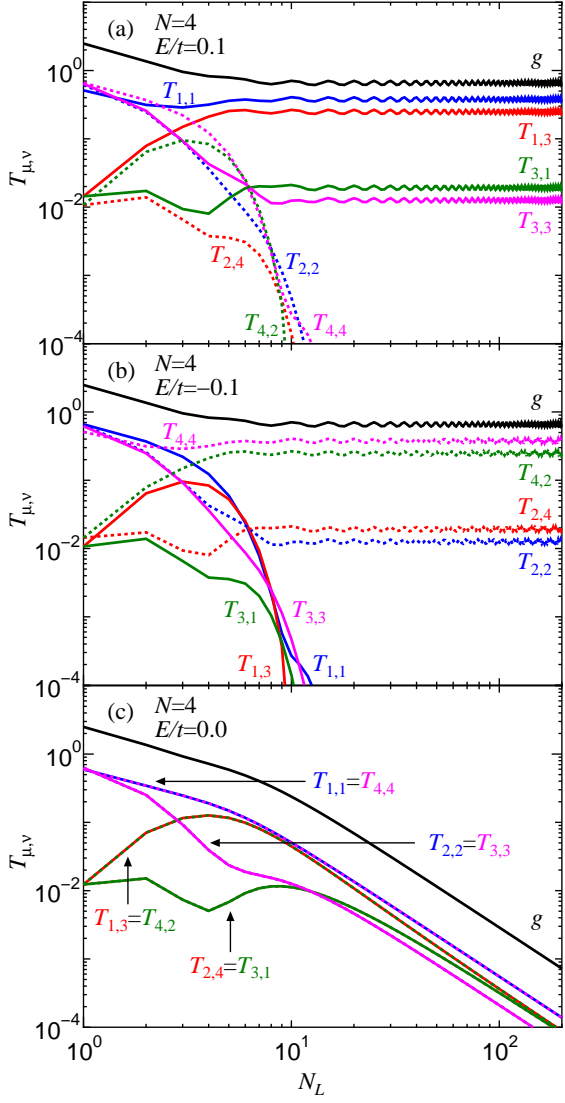} 
\end{center}
\caption{ (Color online)
The transmission probability $T_{\mu,\nu}$ together with the conductance
 $g$ for $N=4$ as a function of $N_L$ 
in the cases of  $E/t = 0.1$ (a), $E/t = - 0.1$ (b) and $E/t = 0.0$ (c). 
}
\label{fig:Tmn}
\end{figure}
\begin{figure}[htb]
\begin{center}
\includegraphics[width=7.9truecm]{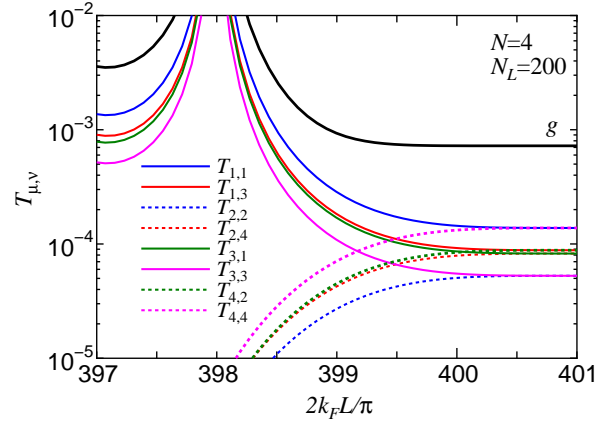} 
\end{center}
\caption{ (Color online)
The transmission probability $T_{\mu,\nu}$ together with the conductance
 $g$ for $N=4$ and $N_L = 200$ as a function of $2 \kf L /\pi$. 
The right end of the horizontal axis $2 \kf L /\pi = 401$ corresponds to
$E/t=0$.  
}
\label{fig:TmnEzero}
\end{figure}
Here, the transmission probabilities $T_{\mu,\nu}$ are discussed in
detail for the $N =$ even cases. 
In these cases, as was already noted, 
the band selective filtering is known to occur
in the transport properties of zigzag GNR
under the external potential uniform in the transverse direction.\cite{Nakabayashi} 
The origin of the filtering is the reflection symmetry in the transverse
direction, which makes the wave function in the direction either even or
odd function.      
Since the present external leads has also reflection symmetry, 
the wave function of the channel defined in the external leads
is either the even or odd function. 
Therefore, the filtering depending on the channel indices is expected 
for the $N=$ even cases.  
We show the transmission probability $T_{\mu,\nu}$ for $N=4$ 
as a function of $N_L$ 
for $E/t = 0.1$ (a),
$E/t= -0.1$ (b) and $E/t=0.0$ (c) in Fig. \ref{fig:Tmn}. 
In the case of $E/t = 0.1$, the transmission probabilities 
$T_{{\rm odd},{\rm odd}}$ becomes finite for $N_L \gg 1$, 
whereas only $T_{{\rm even},{\rm even}}$ survives for $E/t= -0.1$.  
The other transmission processes such as those from the even channel to
the odd one and vice versa are suppressed in the both cases. 
These results can be understood in terms of the parity of the wave
function in the transverse direction; 
The wave function of the channel $\mu$, $u_\mu(l)$ ($l = 1,2,\cdots,N$), 
in the external leads is given as follows, 
\begin{align}
 u_\mu (l) &= \sqrt{\frac{2}{N+1}} \sin k_y(\mu) l, \\
 k_y(\mu) &= \frac{\mu \pi}{N+1}, \quad (\mu = 1,2,\cdots,N)
\end{align} 
and it's parity is positive for $\mu$ = odd, whereas negative for $\mu$
= even.     
On the other hand, in the zigzag GNR with the width $N =$ even, 
among the two bands located around $E/t=0$ 
the wave function for $E/t > 0$ is the even function in the transverse
direction, whereas   
that for $E/t < 0$ is the odd one.\cite{Nakabayashi}
Since the parity of the wave function in the transverse direction 
should be conserved when passing through the zigzag GNR, 
the channels with positive (negative) parity can pass through for $E/t >
0$ ($E/t < 0$). 
At $E/t = 0.0$, the two bands touch with each other and 
two states are degenerate. 
Therefore, the two kinds of channels, \ie from even to even 
and from odd to odd, 
are alive in the case of $E/t=0.0$. 
Actually, the analytical calculation for $N=2$, $T_{1,1}=T_{2,2}\simeq 3/N_L^2$  supports the conclusion shown above.  

Now, we investigate this channel filtering in detail close to  $E/t = 0$ 
where,  as has been already discussed above, 
the anomalous phenomenon that the oscillation of $g$ 
as a function of $\kf L$ disappears is observed. 
In Fig. \ref{fig:TmnEzero}, 
the transmission probabilities 
for $N=4$ and $N_L = 200$ are shown
as a function of $2 \kf L / \pi$ as well as the conductance $g$. 
Here, the $E/t \gsim 0$ case is discussed and the right end of the
horizontal axis corresponds to $E/t=0.0$. 
The last peak in the conductance oscillation is observed at 
$2 \kf L/\pi = 398$ and 
the peak which should be observed at $2 \kf L/\pi = 400$
diminishes. 
In the region, we can see that the transmission processes from the even
channel to the even one, which is prohibited due to parity conservation, 
become possible.  
The regions where the oscillation disappears 
correspond to those with the breakdown of the parity conservation. 
Therefore, the two phenomena can be considered to come from the same
origin. 

Thouless discussed that the energy levels of the finite system attached to the
leads are shifted by the amount of $\hbar/\tau$,  
where $\tau$ is the time it takes for an electron to move to the end of the
system.\cite{Thouless} 
The conductance can be understand from the point of view that 
the states within the shift can pass through the sample.\cite{Altshuler}  
Though he discussed the case with random impurities 
where the motion of the electron is diffusive, 
such a shift of the energy levels in the GNR region of the present system can be expected to occur  
in the presence of the ideal leads. 
As a result, the anomalous phenomena seen in $E/t = 0$ can be expanded
to the $E/t \neq 0$ region. 
However, at present, it is unclear why the conductance at $E/t \simeq 0$ 
shows the discrepancy in the $N_L$ dependences between 
$N =$ even and $N =$ odd case. 
It is likely that the discrepancy in the transport properties 
originates from that in  coupling between the GNR and the normal metal
because the qualitative difference between $N=$ even and odd cannot be observed in the band structure of the isolated GNR.  
Further investigation on the electronic states of GNR attached with the ideal 
leads is necessary in order to  understand the transport properties 
of the normal metal - zigzag GNR - normal metal junctions.     
    
In conclusion, we investigated the transport properties of the 
junction where the zigzag GNR with width $N$ and length $2N_L$ 
is sandwiched by the normal metals expressed by the regular square
lattices. 
The transport property at $E/t=0$ shows different asymptotic behavior 
as the a function of $N_L$, \ie for $N_L \gg 1$ 
the conductance with $N=$ odd approaches to unity 
whereas that in the case of $N =$ even decreases in proportion to
$N_L^{-2}$. 
The conductance at $E/t \neq 0$ shows oscillating behavior as a
function of the energy, which is due to the interference of the electron
wave at the Fermi energy.  
Such a oscillation disappears close to $E/t = 0$. 
In addition, the channel filtering occur in the case of $N =$ even with
the reflection symmetry in the transverse direction. 
The filtering is found also  to show unexpected behavior close to $E/t =
0$.   

\section*{Acknowledgment}
This work was supported by Nara Women's University Intramural Grant for Project Research.


\begin{thebibliography}{99}
\bibitem{Note}
The results for $E/t=0$ has been partly reported in
	Ref. \citen{Yoshioka-Higashibata}.
\end{thebibliography}
\end{document}